\documentstyle[twoside,fleqn,mssl_workshop,psfig]{article}
\newcommand{\etal}{{\it et al.} }
\newcommand{\xmm}{{\it XMM-Newton}}
\newcommand{\chandra}{{\it Chandra}}

\newcommand{\zpup}{\mbox{$\zeta\,$Pup}}
\newcommand{\zori}{\mbox{$\zeta\,$Ori}}

\newcommand{\vinf}{\mbox{$v_\infty$}}

\def\langl{\mathopen{<}}
\def\rangl{\mathopen{>}}

\begin{document}

\title{Quantitative X-ray spectroscopy of massive stars}

\author{L.M. Oskinova$^{\rm a}$, W.-R. Hamann$^{\rm a}$, A. Feldmeier
\address{Potsdam University, Am Neuen Palais 10, Potsdam, 14469 Germany}}

\begin{abstract}
Radiative transfer in a clumped winds is used to describe X-ray emission 
line profiles observed in the \xmm\ RGS spectrum of the OI star $\zeta$~Puppis.
It is shown that this X-ray spectrum can be explained as originating from a
multi-temperature collisional plasma located in the wind acceleration zone. 
The  X-rays are attenuated in the clumped stellar wind, which gives 
characteristic profiles to the emergent lines. We specifically study  
the N\,{\sc vii} emission line in the spectrum of \zpup. Long RGS exposures 
reveal that the N\,{\sc vii} line profile is structured. On the basis of 
our \zpup\ atmosphere model, we rule out the presence of N\,{\sc vii} 
in the cool wind component. We suggest that the detailed N\,{\sc vii} 
line structure is due to self-absorption in the hot plasma.
Wind clumping also affects the transfer of ionizing radiation in high-mass 
X-ray binaries (HMXBs). We derive analytical formulae for the ionisation 
parameter in dependence on the parameters of wind clumping. 
\end{abstract}

\maketitle

\section{Clumping of the stellar wind}

%
\begin{figure}[t] 
\centerline{\psfig{file=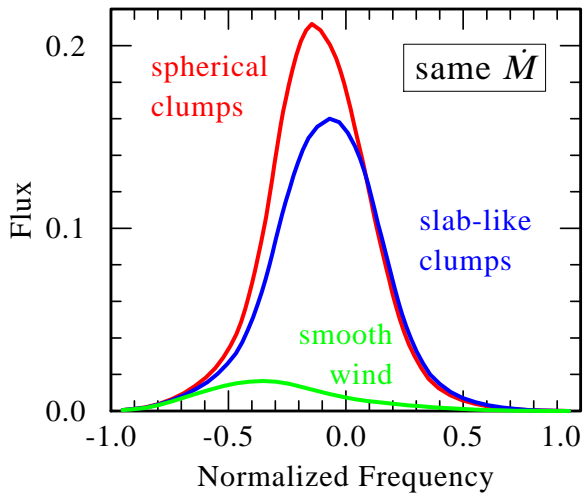,width=3.0in}}
\caption{Model X-ray emission line profiles. The same stellar 
parameters were assumed for all three profiles, except of different 
clumping  properties as indicated.}
\label{fig:sfh}
\end{figure}
%

Stellar winds of O-type stars are strongly clumped as confirmed by the 
growing number of observational evidences. 
The observed variability in the  He\,{\sc ii}~4686\,\AA~emission line in 
\zpup~was explained as an excess emission from the  wind clumps \cite{ev98}.
The H$\alpha$ variability in a large sample of O stars was reproduced by a 
wind model consisting of coherent shells \cite{mar05}. 

\begin{figure}[t] 
\centerline{\psfig{file=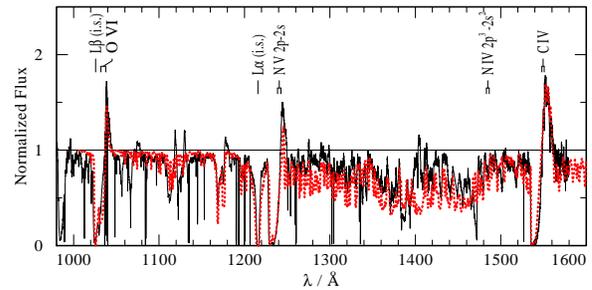,width=3.0in}}
\caption{The EUV spectrum of \zpup, observed with FUSE (thin line),
compared to a PoWR model spectrum (dotted). The resonance doublets of
C\,{\sc iv}, N\,{\sc v} and O\,{\sc vi} are well reproduced, as well as
the ``forest'' of iron-group lines. The O\,{\sc vi} doublet can only be
fitted with models when assuming that a diffuse X-ray field causes
additional ionisation. }
\label{fig:zpmod}
\end{figure}        
%
The empirical mass-loss estimates shall be generally reduced by factors 
of a few  in clumped winds \cite{hk98}. From a study of FUV spectra of 
40 Galactic O-stars it was concluded that their mass-loss rates should 
be reduced by up to one order of magnitude due to wind clumping 
\cite{ful05}. An analysis of the X-ray emission line profiles from 
$\delta$~Ori, \zpup, and \zori~ \cite{mil02,kr03,cohen06} has shown 
that the attenuation by the stellar wind is significantly smaller 
than expected from homogeneous wind models. The observed attenuation 
of X-rays in the massive binary $\gamma^2$\,Vel is much weaker than 
expected from smooth stellar wind models, implying strong wind 
clumping \cite{sch04}. Similar conclusions were reached from the 
X-ray observations of WR\,140  \cite{pol05}.

\begin{figure*}[t] 
\vspace{10pt}
{\psfig{file=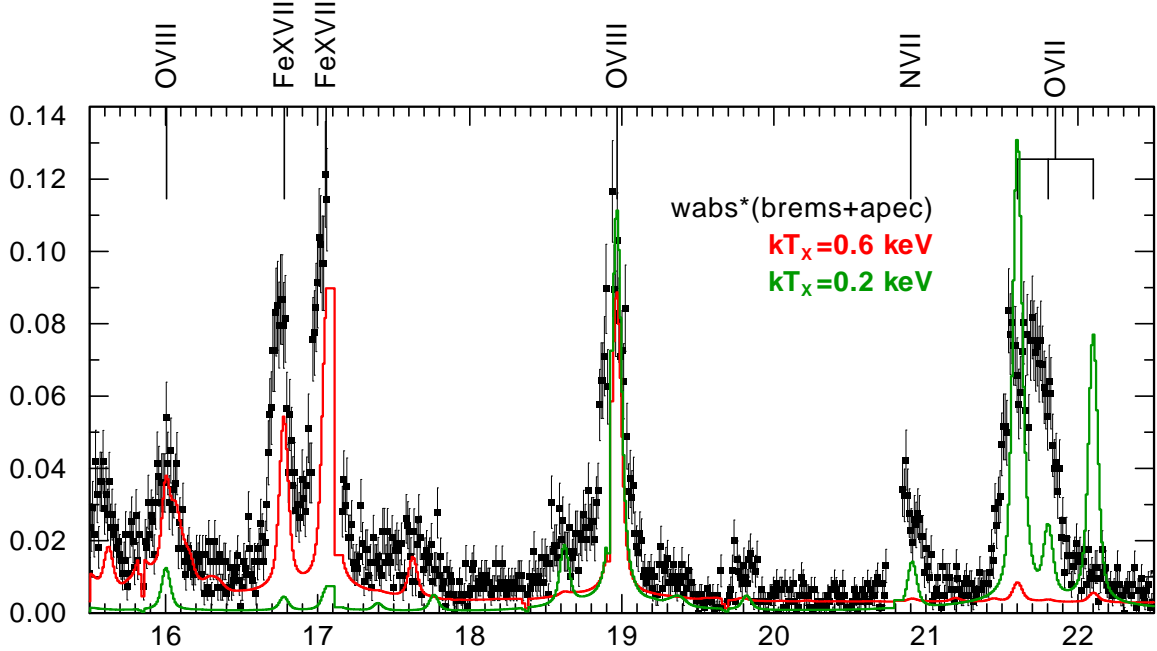,width=6.0in,angle=-90}}
\caption{Part of \xmm RGS spectrum of \zpup. Overplotted are two 
{\em wabs$\times$apec} models with different temperatures. The level of 
continuum and the line ratios are in agreement with a plasma in collisional 
equilibrium.}
\label{fig:O8}
\end{figure*}

The X-ray emission line spectra of several wind-fed high-mass X-ray 
binaries (HMXB) can be explained  as originating in a wind where cool 
dense clumps are embedded in rarefied photoionized gas \cite{sako03}. 
The stochastic variability of X-ray light-curve  observed by \xmm\  
in 4U~1700-37 were explained as the result of the neutron star feeding 
by a strongly clumped  stellar wind \cite{vdm05}. 

Clumping of the stellar wind is thought to be a consequence of the 
intrinsic instability of the radiative-driving mechanism \cite{Lucy70}.
The 1D hydrodynamic modeling  predicts that line-driven stellar winds 
are strongly inhomogeneous, starting from close to the photosphere 
\cite{AF97}. Strong gas compression leads to the formation of cool 
clumps. The space between clumps is essentially void, but small gas
parcels are ablated from the clumps and accelerated to high speed. 
When a parcel catches up and rams into the next outer clump, the gas 
is heated and emits thermal X-rays. The X-rays propagate trough the 
wind and suffer absorption by the cool clumped component of the wind.  
gets. Thus the wind consists of two disjunctive components: hot 
($\sim 10^7$\,K) gas parcels emitting X-rays and compressed cool 
($\sim 10^5$\,K) fragments that attenuate this radiation. 

\begin{figure*}[t] 
\vspace{10pt}
{\psfig{file=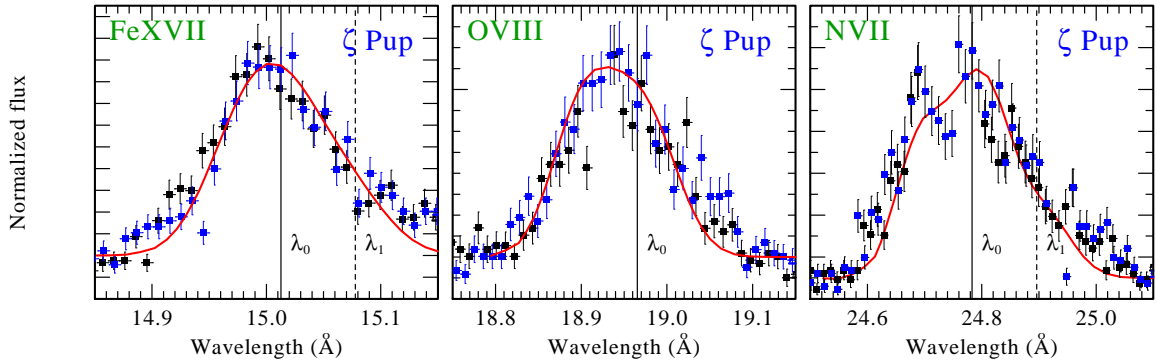,width=6.0in,angle=-90}}
\caption{Lines observed in \zpup~ (overplotted are RGS1 and RGS2). 
The solid lines are the clumped wind model with 
$\dot{M}=8.8\times 10^{-6}\,M_\odot\,{\rm yr}^{-1}$. 
The cool-wind opacity is calculated from a PoWR model. The rest-frame 
wavelength is denoted $\lambda_0$. The rest-frame wavelengths of blended 
lines are denoted $\lambda_1$.}
\label{fig:ln}
\end{figure*}


Radiative transfer in a clumped wind is different from transfer in a 
homogeneous wind \cite{pom91}. Formation of X-ray emission lines in  
clumped winds of massive stars was extensively studied in 
\cite{feld03,osk04,osk06}. The main findings of these studies are

\bigskip
\noindent 
- The opacity in a clumped wind is reduced compared to a 
homogeneous wind of the same mass-loss rate.

\medskip\noindent
- The  wavelength-dependence of the opacity is weaker in clumped 
winds; if clumps are optically thick, the opacity becomes grey.

\medskip\noindent
- X-ray emission line profiles are nearly symmetric and have 
similar shapes across the spectrum.

\medskip\noindent
- The line profiles observed in X-ray spectra of single O stars 
can be reproduced in the framework of a clumped wind model.
\bigskip

Fig.~\ref{fig:sfh} shows model line profiles for smooth and clumped
winds. Two models of clumped wind are shown: the wind which consists 
of the spherical and of the flattened clumps. The line flux in the 
Fig.~\ref{fig:sfh} is normalized such that it is unity for a line which 
is not affected by absorption. As can be seen from Fig.~\ref{fig:sfh} 
the emergent flux in the line is lower in a smooth wind that in a   
clumped wind model. It is assumed that the total number of clumps 
in the wind is constant. Clumps form at random times near some inner 
boundary and propagate radially according to a monotonically increasing 
velocity law. An average time separation between clumps is equal to the 
wind flow time  $T_{\rm fl}=R_*/v_\infty \sim 1$\,hr. The line profiles 
are affected by the assumed clump geometry. A model with spherical clumps 
assumes that the optical depth across a clump has no angular dependence 
(see \cite{oc06}). Slab-like clump models assume clumps, that are 
geometrically thin and aligned. In the latter model the optical depth 
across the clump is angular-dependent (see \cite{feld03,osk04,osk06}) 
and the resulting line profile is more symmetric. Future 2D  hydrodynamic 
models will shed new light on the clump geometry.  

\section{Stellar atmosphere model}

While our 2D clumped-wind model is used to calculate the formal radiative 
transfer, the opacity of the cool medium is obtained with the
Potsdam Wolf-Rayet (PoWR) non-LTE code 
\footnote{www.astro.physik.uni-potsdam.de/$\sim$wrh/PoWR/powrgrid1.html}.
Shown in Fig.\,{\ref{fig:zpmod}} is the observed FUSE spectrum of \zpup\ 
with overplotted model. The model and observations are flux-calibrated. 
An X-ray field of the observed level has been included in the model to fit 
the observations.  As can be seen in Fig.\,{\ref{fig:zpmod}} the P~Cyg line 
profile of O\,{\sc vi} is nicely reproduced by the model. It was the presence 
of such high ions as O\,{\sc vi} observed in O-star spectra, which led 
Cassinelli \& Olson \cite{cas79} to postulate the presence of X-rays 
in stellar winds, prior to their actual discovery. In agreement with 
this classical work, our state-of-the-art model cannot reproduce the 
observed O\,{\sc vi} line unless in the presence of X-rays. 
 
The scattering by the resonance O\,{\sc vi} line in the wind is needed to be 
taken into account for the analysis of the $\cal R$ ratio between forbidden 
and intercombination lines in helium-like Mg\,{\sc xi} \cite{fir06}. The  
transition between $2^3S_1\rightarrow 2^3P_1$ levels at 
$\lambda\,1034.31$\,\AA \ is between O\,{\sc vi} doublet at $\lambda\,1031.91$
\,\AA \ and $\lambda\,1037.61$\,\AA. Thus, the full radiative transfer 
is needed to evaluate the effect of radiation field on the lines ratio in 
helium-like ions. This is fully accounted for in our atmosphere models. 

From the line ratios in helium-like ions observed by \chandra\ and \xmm\ 
in \zpup \ we infer the X-ray formation region to be located approximately 
between $1.5\,R_*$ and $10\,R_*$ \cite{osk06}. This is consistent with the 
radii of formation obtained independently from line profile fits. Detailed 
analysis of the forbidden-to-intercombination line ratios in 
\zpup, \zori, $\iota$\,Ori, and $\delta$\,Ori is presented 
in \cite{fir06}. 
\begin{figure}[t] 
\centerline{\psfig{file=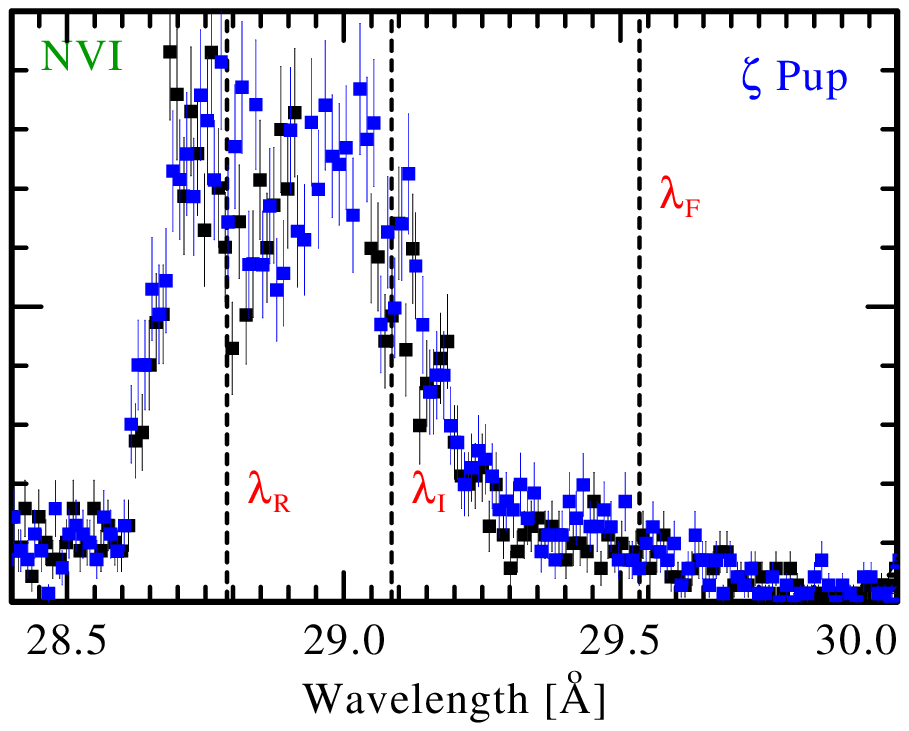,width=2.6in}}
\caption{Observed N\,{\sc vi} line in spectrum of \zpup.}
\label{fig:nvi}
\end{figure}    

\section{Analysis of \xmm\ RGS spectrum of \zpup}

We have analyzed the  archival RGS spectra of \zpup.
General properties of \zpup's X-ray spectrum are consistent with 
a multi-temperature plasma in collisional equilibrium (CIE), as 
confirmed by a study of differential emission measure distribution 
\cite{woj05}. Fig.~\ref{fig:O8} shows part of the RGS spectrum with 
overplotted {\em wabs*apec} CIE models for two temperatures. As can 
be seen, the ratio of O\,{\sc viii} 
Ly$\beta(\lambda 16.20)/{\rm Ly}\alpha(\lambda 18.99)$ is consistent with 
a plasma temperature of $kT_{\rm X}\approx 0.6$\,keV. On the other hand, 
the strength of O\,{\sc vii} helium-like line indicates presence of a
$kT_{\rm X}\approx 0.2$\,keV plasma. Importantly, the level of continuum 
seen in this high quality spectrum is in agreement with the inferred plasma 
temperatures. 

Continuum was only marginally detected in previous less sensitive 
observations \cite{kahn01} and its alleged weakness was used as an argument 
to support the idea, that the shocks in  O star winds are collisionless  
(Rauw, Pollock \& Naz{\'e}, these proceedings). A recent study of 
collisionless damping in stellar winds shows that in rotating magnetized 
stars  this mechanism can indeed be important \cite{suz06}. However, 
from both dynamic and energetic arguments it was demonstrated that it 
can operate effectively only in stars with spectral type B3V or later, 
but not in O stars \cite{suz06}. 

\begin{figure}[!] 
\centerline{\psfig{file=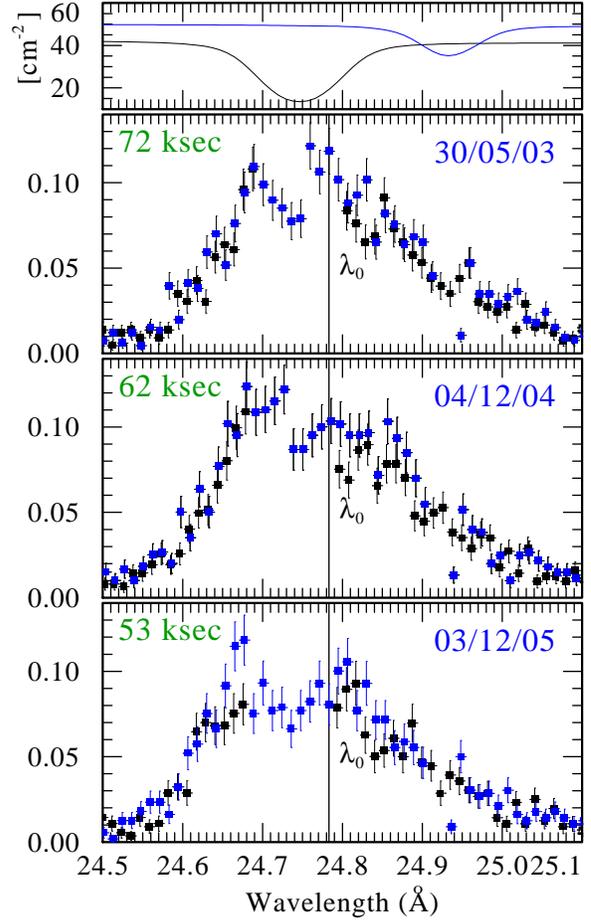,width=3.0in}}
\caption{Observed N\,{\sc vii} lines in spectrum of \zpup\ 
(RGS1 and RGS2 are overplotted). The upper panel shows the sensitivity 
areas of RGS1 and RGS2 across the line. The exposure time and the date of 
observation are in left and right upper corners.}
\label{fig:nvii}
\end{figure}  

In order to fit the  observed line profiles we have added the observed level 
of continuum to the model lines. Fig.\,\ref{fig:ln} shows  the most 
prominent emission lines in the RGS spectrum. The model lines were obtained
with our 2D clumped wind model \cite{osk04}.  The region of X-ray line 
formation is constrained from an analysis of the {\em fir} line ratios 
in He-like ions \cite{osk06,fir06}. Therefore, the only free parameter in the  
modeling is the {\em fragmentation frequency}. Our model assumes time 
averaged stationary mass-loss rate and the average number of clumps in 
the wind at a given moment of time is constant. The fragmentation frequency
is an inverse time interval at which moving clumps are crossing some 
arbitrary reference radius in the wind \cite{osk06}. We prefer to use this 
parameter instead of {\em porosity length} \cite{oc06}, since the latter 
is radius-dependent whereas the former is not.  

The model lines shown in Fig.\,\ref{fig:ln} are obtained  using  mass-loss 
estimate which does not account for wind clumping \cite{Rep04}. The 
fragmentation frequency for all models is $10^{-4}\,{\rm s}^{-1}$. 
This corresponds to the inverse flow time, $T_{\rm fl}=\vinf/R_*$, in 
the wind of \zpup. Using more realistic mass-loss rates that are smaller 
by a factor of few will result in slightly higher fragmentation frequency. 
The Fe\,{\sc xvii} ($\lambda 15.01$) line is wide and cannot be 
satisfactorily fitted unless the line formation region extends further out 
in the wind.  We assumed here that the line is blended with Fe\,{\sc xix}. 
The emissivity of Fe\,{\sc xix}\, 
($\lambda 15.08$) is about 30\% of Fe\,{\sc xvii}\,($\lambda 15.01$), 
assuming a temperature $kT_{\rm X}\sim 0.7$\,keV. In the most red part 
of the wing of Fe\,{\sc xvii}, the emission from 
O\,{\sc viii} ($\lambda 15.17$) is also likely to contribute.

\begin{figure}[t] 
\centerline{\psfig{file=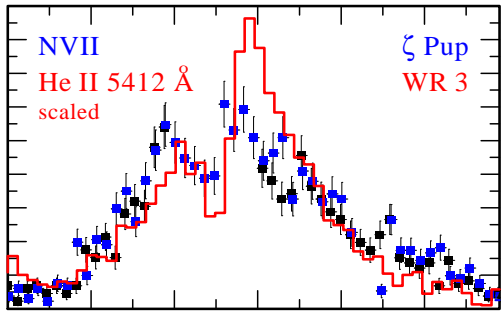,width=2.6in}}
\caption{Comparison between the shape of the optically thick He\,{\sc ii} 
line observed in WR\,3 and the N\,{\sc vii} in the X-ray spectrum of \zpup.}
\label{fig:wr3}
\end{figure}    

The O\,{\sc viii} ($\lambda 18.97$) is easily fitted with 
the present model of a clumped wind.

Several authors have noted that the profile of N\,{\sc vii} 
($\lambda 24.785$) is different from that of other lines 
\cite{kahn01,cas01}. It was recently suggested  that 
this line is blended with N\,{\sc vi} ($\lambda 24.898$) \cite{pol06}.
Following this suggestion we model  N\,{\sc vii} line as a blend. 
Assuming ionisation equilibrium, the ratio of N\,{\sc vi} to N\,{\sc vii} 
is sensitive to the temperature. Fig.\,\ref{fig:nvi} shows the observed 
N\,{\sc vi} triplet line. The measured 
ratio ${\cal G} ={\it f+i}/{\it r}= 1.1\pm 0.1$
corresponds to a temperature $T_{\rm X}\approx 1$\,MK \cite{por01}. At 
this temperature, the  emissivity of  N\,{\sc vi}\,($\lambda 24.898$) 
is half of the emissivity of N\,{\sc vii}\,($\lambda 24.785$) 
(ATOMDB v.1.3.1). Using these values in  clumped wind model, the 
resulting fit satisfactorily reproduces the N\,{\sc vii} line width, 
as is shown in  Fig.\,\ref{fig:ln}. 

Figure\,\ref{fig:nvii} shows three different \xmm \ observations of 
the N\,{\sc vii} line and the sensitivity areas of  RGS1 and RGS2. 
The ``ragged`` structure of the line profile and the dip at 
$\lambda 24.74$\AA \ is clearly seen in all data sets. While the depth 
of the dip is variable, its location is relatively constant.   

We employed PoWR model to study whether the absorption in the cool wind 
can cause the observed structure of the N\,{\sc vii} line. N\,{\sc vii} 
can be produced in the cool wind via Auger ionisation of N\,{\sc v}. 
However, our atmosphere models show that N\,{\sc vii} is absent in the 
cool wind, even when assuming that the X-ray field is significantly 
stronger in \zpup \ wind than observed.

The line structure seen in N\,{\sc vii} is typical for the optically  
thick lines in the winds of hot stars. Fig.\,\ref{fig:wr3} shows the 
optically thick He\,{\sc ii} line in the spectrum of WR\,3, overplotted 
with the N\,{\sc vii} line in \zpup's  spectrum. The optical depths in 
the resonance lines of the most abundant ions in the hot X-ray emitting 
component of stellar winds can be larger than unity \cite{ig02}. Nitrogen 
is overabundant in  \zpup, while oxygen and carbon are depleted. We find 
that the optical depth in N\,{\sc vii} line can be large enough 
to result in self-absorption.

\section{The photoionisation parameter in the clumped wind of HMXBs}

Clumping of the stellar wind has a profound effect on X-ray 
emission from HMXBs \cite{sako03}. We studied the influence of clumping
on photoionised regions. In the following we consider the case of 
optically thick clumps. The same notations as in \cite{hmg77} are used. 
We assume that the wind consists of two components: dense clumps and 
a rarefied interclump medium. Applying the continuity equation we define 
the mean particle number density as
\begin{equation}
n(r_*)  \equiv \frac{\dot{M}}{4\pi\mu m_{\rm H} v(r_*)r_*^2},
\label{eq:avn} 
\end{equation}
\noindent where $\mu$ is mean molecular weight and $m_{\rm H}$ is the 
atomic mass unit.  If the density enhancement in clumps is 
$D=\langl n^2\rangl/\langl n \rangl^2$, and the volume filling factor of 
clumps is $f_{\rm V}$ then the particle number density of the interclump 
gas is
\begin{equation}
n_{\rm ic}=(1-Df_{\rm V}) n 
\end{equation}

Assume for simplicity that the interclump gas is optically thin and the 
velocity is constant. Correcting the mean radiation intensity for the 
absorption in the clumps
\begin{equation}
J_{\nu}=\frac{L_{\rm X}}{4\pi r_*^2}\cdot f(kT)\cdot 
{\rm e}^{-\bar{\tau}(r_*)},
\label{eq:jnu}
\end{equation}
where $f(kT)$ is a function of atomic parameters and temperature, and 
$\bar{\tau}$ is an effective optical depth. When all clumps are
opaque ($\tau_{\rm clump}\gg 1$), the effective optical depth 
reflects  geometrical distribution of the clumps, and is found to be grey 
\cite{feld03,osk04}

\begin{equation}
\bar{\tau}=\frac{n_0r_*}{v}~~~~~~~~~({\rm if}~~\tau_{\rm clump}\gg 1),
\label{efft}
\end{equation}
where $n_0$ is the fragmentation frequency. Using the ionisation balance 
equation, the ionisation parameter in the interclump medium is
\begin{equation}
\xi=\frac{L_{\rm X}}{\langle n \rangle r_*^2}\cdot
\frac{{\rm e}^{-\bar{\tau}}}{1-Df_{\rm V}}.
\label{eq:ksi}
\end{equation}
For a very high fragmentation frequency, which recovers the case of 
a smooth wind  $Df_{\rm V}\approx 0$, Eq.\,\ref{eq:ksi} reduces to the 
usual definition of $\xi$. Parameter $q$ \cite{hmg77} in a clumped wind   
can be written as
\begin{equation}
q= \frac{r_*^2}{r_{\rm X}^2}\cdot {\rm e^{-\bar{\tau}}}.
\label{eq:q}
\end{equation}

We conclude that clumping affects both the ionisation balance and the 
size of the photoionized region.A higher ionisation and a larger size 
of the photoionized region can be expected.     


\normalsize

\section*{ACKNOWLEDGEMENTS}
Based on observations obtained with \xmm, an ESA science mission with 
instruments and contributions directly funded by ESA Member States and NASA.
LMO and AF acknowledge support from  DFG grant Fe 573/2-1.                   
           

\small
\begin{thebibliography}{9}

\bibitem{ev98}
Eversberg, T., L{\`e}pine, S., Moffat, A.F.J . 1998, ApJ, 494,799

\bibitem{mar05}
Markova, N., Puls, J., Scuderi, S., Markov, H. 2005, A\&A, 440, 1133

\bibitem{hk98}
Hamann, W.-R. \& Koesterke, L. 1998, A\&A, 335, 1003

\bibitem{ful05}
Fullerton, A.W., Massa, D.L., Prinja, R.K. 2006, ApJ, 637, 1025

\bibitem{mil02}
Miller, N.A., Cassinelli, J.P., Waldron, W.L., \etal 2002, ApJ, 577, 951

\bibitem{kr03}
Kramer, R.H., Cohen, D.H., Owocki, S.P. 2003, ApJ, 592, 532

\bibitem{cohen06}
Cohen, D., Leutenegger, M.A., Grizzard, K.T., \etal 2006, MNRAS, in press, 
astro-ph/0602599

\bibitem{sch04}
Schild, H. \etal 2004, A\&A, 422, 177 

\bibitem{pol05}
Pollock, A.M.T., Corcoran, M.F., Stevens, I.R. \etal 2005, ApJ, 629, 482

\bibitem{sako03}
Sako M., Kahn S.M., Paerels F., \etal 2003, in High Resolution X-ray 
Spectroscopy with {\em XMM-Newton} and {\em Chandra}, ed. G. Branduari-Raymont 
(astro-ph/p309503)  

\bibitem{vdm05}
van~der~Meer, A., Kaper, L., Di~Salvo, T., \etal 2005, A\&A, 432, 999

\bibitem{Lucy70}
Lucy L.B. \& Solomon P.M., 1970, ApJ, 159, 879

\bibitem{AF97} 
Feldmeier, A., Puls, J., Pauldrach, A.W.A. 1997, A\&A, 322, 878

\bibitem{pom91}
Pomraning G.C., 1991, Linear kinetic theory and particle transport in 
stochastic mixtures, Singapore; New Jersey: World Scientific. Series on 
advances in mathematics for applied sciences 7

\bibitem{feld03} 
Feldmeier, A., Oskinova, L., Hamann, W.-R. 2003, A\&A, 403, 217

\bibitem{osk04}
Oskinova, L.M., Feldmeier, A., Hamann, W.-R. 2004, A\&A, 422, 675

\bibitem{osk06}
Oskinova, L.M., Hamann, W.-R.,  Feldmeier, A. 2006, MNRAS, submitted, 
astro-ph/0603286 

\bibitem{oc06}
Owocki S.P. \& Cohen D.H., 2006, ApJ, in press, astro-ph/0602054 

\bibitem{cas79}
Cassinelli, J.P. \& Olson, G.L. 1979, ApJ, 229, 304

\bibitem{fir06}
Leutenegger, M.A., Paerels, F.B.S., Kahn, S.M., Cohen, D.H., 2006, ApJ, 
submitted  
 
\bibitem{woj05}
Wojdowski, P.S. \& Schulz, N.S. 2005, ApJ, 627, 953

\bibitem{kahn01}
Kahn S.M., Leutenegger M.A., Cotam J., \etal 2001, A\&A, 276, 117

\bibitem{suz06}
Suzuki, T.K., Yan, H., Lazarian, A., Cassinelli, J.P. 2006, ApJ, 640, 1005

\bibitem{Rep04}
Repolust, T., Puls, J., Herrero, A., 2004, A\&A, 415, 349

\bibitem{cas01}
Cassinelli J.P., Miller N.A., Waldron W.L., \etal 
2001, ApJ, 554, L55

\bibitem{pol06}
Pollock, A.M.T. \& Raassen, A.J.J. 2006, A\&A, submitted

\bibitem{por01}
Porquet D., Mewe R., Dubau J., \etal 2001, A\&A, 376, 1113

\bibitem{ig02}
Ignace, R. \& Gayley, K.G. 2002, ApJ, 568, 954

\bibitem{hmg77}
Hatchett, S. \& McGray, R. 1977, ApJ, 211, 552


\end{thebibliography}
\end{document}